\documentclass[twocolumn]{aastex62}

\newcommand\Hlim{H_{lim}}
\newcommand{\au}{{\rm au}}

\usepackage{soul}

\usepackage{outlines}
\usepackage{enumitem}
\setenumerate[1]{label=\Roman*.}
\setenumerate[2]{label=\Alph*.}
\setenumerate[3]{label=\roman*.}
\setenumerate[4]{label=\alph*.}

\definecolor{darkbrown}{RGB}{100,50,30}

\shorttitle{Empirical Completion Limit Model}
\shortauthors{Hendler \& Malhotra}

\usepackage[section]{placeins}

\begin{document}

\title{Observational Completion Limit of Minor Planets from the Asteroid Belt to Jupiter Trojans}

\correspondingauthor{Nathanial P. Hendler}
\email{equant@lpl.arizona.edu}

\author[0000-0002-3164-0428]{Nathanial P. Hendler}
\affil{Lunar and Planetary Laboratory, The University of Arizona, Tucson, AZ 85721, USA}

\author[0000-0002-1226-3305]{Renu Malhotra}
\affil{Lunar and Planetary Laboratory, The University of Arizona, Tucson, AZ 85721, USA}



\begin{abstract}

With the growing numbers of asteroids being discovered, 
identifying an observationally complete sample is essential for statistical analyses and for informing theoretical models of the dynamical evolution of the solar system.
We present an easily implemented method of estimating the  
empirical observational completeness in absolute magnitude, $\Hlim$, as a function of semi-major axis.
Our method requires 
fewer assumptions and decisions to be made in its application, making results more transportable and reproducible amongst studies that implement it, as well as scalable to much larger datasets of asteroids expected in the next decade with the Vera C.~Rubin Observatory's Legacy Survey of Space and Time (LSST). 
    Using the values of $\Hlim(a)$ determined at high resolution in semimajor axis, $a$, we demonstrate that the observationally complete sample size of the main belt asteroids is larger by more than a factor of 2 compared to using a conservative single value of $\Hlim$, an approach often adopted in previous studies. Additionally, by fitting a simple, physically motivated model of $\Hlim(a)$ to $\sim 7 \times 10^5$ objects in the Minor Planet Database, our model reveals statistically significant deviations between the main belt and the asteroid populations beyond the main belt (Hungarias, Hildas and Trojans), suggesting potential demographic differences, such as in their size, eccentricity or inclination distributions.

\end{abstract}


\section{Introduction} \label{sec:intro}

The distributions of orbital elements and the physical characteristics of the
minor planets between Mars and Jupiter are of great scientific interest
because they are thought to retain important clues about the formation of the terrestrial planets and the origin and
evolution of the Solar System.  As of September 2019, more than eight hundred
thousand of these minor planets have been discovered, offering a
statistically large sample to permit deep analyses for dynamical structures and
their correlations with their physical properties.  However the discovered
sample is subject to observational biases that must be understood for the
proper interpretation of their demographics.  Because these minor bodies are
usually detected by their reflected sunlight, the observed sample
suffers from the selection effect that brighter objects are more easily
detectable than fainter objects.  This bias manifests as an artificial peak in
the distribution of the absolute magnitude, $H$, even when the intrinsic
population of fainter/smaller objects grows rapidly with $H$.  Therefore, in
attempting demographic statistical analyses, it is often important to identify
an observationally complete sample.

In the previous literature, the absolute magnitude at which a sample becomes
observationally incomplete, $\Hlim$, has typically been determined using one of
two methods.  The first, and not uncommon method is for a single value to be
given without explicit justification, or to be sourced from an older work.
The second approach, which we refer to as the
exponential-fitting method, makes the assumption that the intrinsic
population's size distribution is expected to increase as a power-law as one
considers objects of smaller and smaller diameters.  Assuming constant albedo,
$H$ then follows an exponential function, and the $H$ value at which a decline
in the population of observed objects is seen relative to a fitted exponential
function is taken to be $\Hlim$.  Use of the exponential-fitting
method requires several assumptions.  First, it must be assumed that the
asteroids follow a single power-law size distribution. Second, a range of $H$
must be chosen to apply a linear regression.  Third, the magnitude at which the
deviation from linear is extreme enough to define $\Hlim$ must be chosen.  The
latter two require subjective decision-making.  Alternatively, the peak of the
$H$ distribution may be taken as the completeness limit
\citep[e.g.][]{Ryan2015}, a method that we adopt and expand in the present
work.

Additionally, a common approach in previous studies has been to adopt a single value of $\Hlim$ for the entire asteroid belt or for large subsets, such as the inner, middle and outer belts.
For example,
\cite{Gladman2009} and \cite{Malhotra2017} adopt $\Hlim = 15$, and
\cite{Michtchenko2016} and \cite{2018AJ....155..143C} adopt $\Hlim = 15.5$. 
\cite{Dermott2018} adopted $\Hlim=16.5$, while noting that the observational completion limit within the inner main belt varies from $\Hlim=17$ at 2.1~au to $\Hlim=16.5$ at 2.5~au. 

In order to better
assess the demographics of the asteroids at higher resolution in semi-major
axis we have found a need to model the completion limit as a function of
semi-major axis, $\Hlim(a)$, and to ascertain that its value is contemporary with
our sample population because $\Hlim(a)$ evolves with time as observational surveys
go deeper with technological advances.

We have developed a method to measure $\Hlim(a)$ from observations, and
to describe it with a simple, physically motivated, parametric model.  We think that our
approach will provide researchers with completion limit models for the 
asteroids and Jupiter Trojans that can be used to more accurately and
consistently simulate observations, debias observations for statistical
analyses, and model/synthesize minor-planet populations.  This approach will
lead to more reproducible  
results, and allow for the completion
limit to be updated consistently over time as more minor bodies are discovered. 

\section{Asteroid Data} \label{sec_data}

For our analysis, we use the Minor Planet Center Orbit Database
(MPCORB\footnote{https://www.minorplanetcenter.net/iau/MPCORB Downloaded on
2019-09-29}).  The MPCORB database provides, among other things, the osculating
orbital elements of observed Solar System minor bodies.  As of September 29th
2019, this database contains data for 845,049 minor planets of which 729,567
have had multiple observations near opposition, and therefore have orbital elements and absolute
magnitudes that are reliably known.  In this work, we utilize two parameters from
MPCORB: semi-major axis and absolute magnitude ($a$, $H$).

We investigated using proper elements from the AstDyS database \citep{Knezevic2000}, and
decided that (a) because the present day orbital parameters are more appropriate
for understanding observational bias, and (b) because AstDyS does not provide proper elements for all objects, the MPCORB database was more appropriate for
this work.

\section{Analysis} \label{sec_analysis}

\subsection{$\Hlim$ in radial bins} \label{sec_analysis_bins}

\begin{figure*}
\plotone{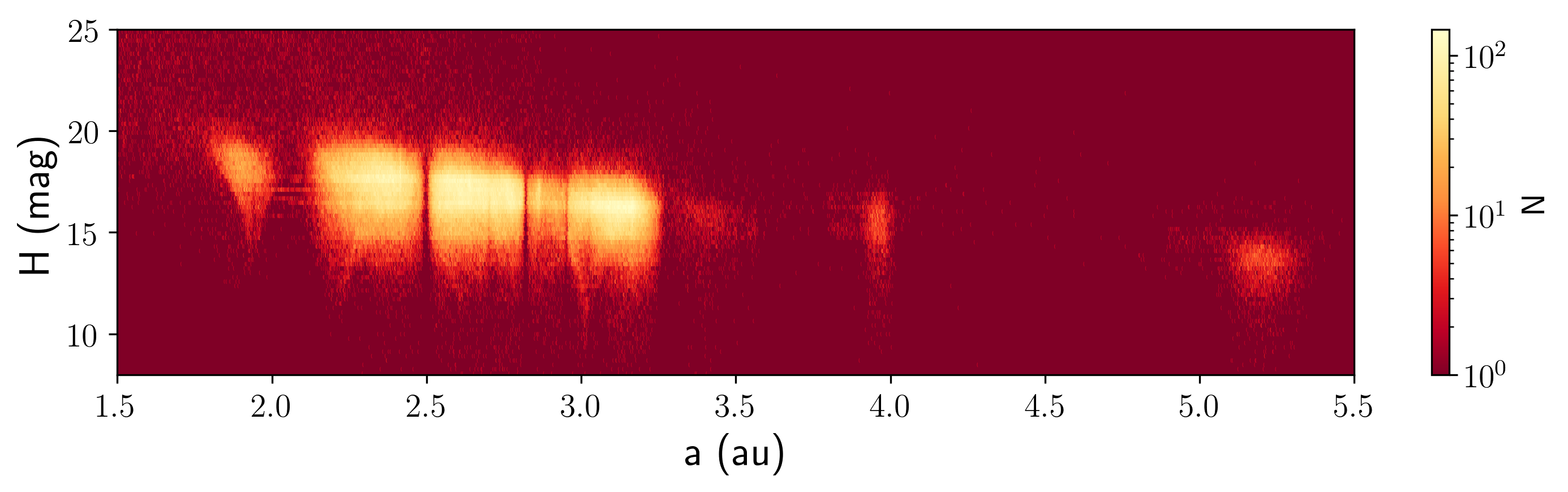}
\caption{
Heatmap of observed H magnitude for minor planets with a semi-major axis
    between 1.5 and 5.5~au.  This range of semi-major axes includes the
    Hungarias, main belt asteroids, Hildas and Trojans.  The regions of
    greatest population density are shown in yellow.  The completion limit
    as a function of semi-major axis, $a$, is evident by eye as a downward
    sloping trend in the upper envelope of the highest density regions.
\label{fig_heatmap}
}
\end{figure*}

The upper panel of Figure~\ref{fig_heatmap} is a 2D density plot (i.e., heatmap
or 2D histogram) showing the distribution of absolute magnitude ($H$) as a
function of semi-major axis, $a$, in the range 1.5--5.5~au, for objects
reported by the Minor Planet Center MPCORB database of September 29, 2019.  In
this figure, a general trend can be seen where the most populated $H$ at a
given distance decreases with $a$. We argue below that a sepcific functional form of this trend is expected
due to the flux limited completion limit of minor planet observations. 

From this sample, we empirically determine the completion limit as a
function of semi-major axis.  We measure the empirical $\Hlim$ at
each radial location by binning objects in both $a$ and $H$ and then
identifying the $H$ bin with the greatest number of objects,
the center of which we then take to be the value of $\Hlim$
at that radial distance.
(The measurement uncertainty on $\Hlim$ is discussed in Section \ref{sec_model_fitting}.)
This
method is a generalization of the exponential-fitting method.
However, our method requires fewer assumptions and fewer subjective
decisions to be made in its application, making results more
transportable and comparable between studies that implement it.

\begin{figure}
    \plotone{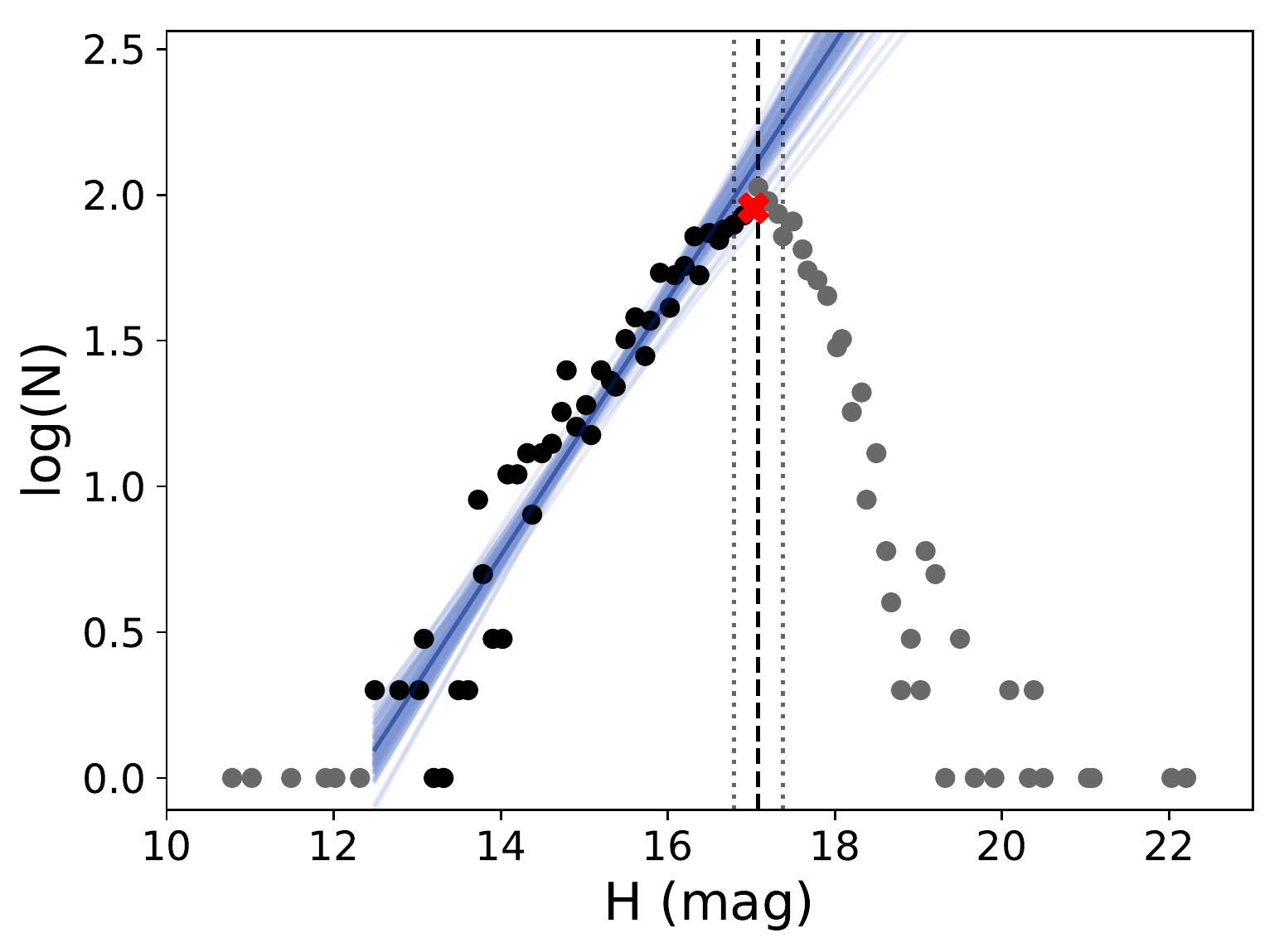}
    \caption{
    Illustration of our method of determining $\Hlim$ for a single
    semi-major axis bin, and its comparison with the
    exponential-fitting method commonly used in previous works.  In this example, the semi-major axis bin is $a \in (2.693, 2.695]$~(au), and the $H$
    distribution is shown as the black and grey dots.  The most populated $H$
    bin (denoted by the dashed vertical line at $H=17.1$) is taken to be
    the value of $\Hlim$ for this semi-major axis bin.  The estimated
    confidence interval for $\Hlim$ is shown with vertical dotted 
    lines.  For the exponential-fitting method, data between H=12.5 and H=16.75
     (black dots) is fit with linear regression; the best-fit
    is shown as the dark blue line and the related confidence interval is shown in
    lighter blue around it.  This figure, including the point marked with a red 'x', is further
    discussed in Section \ref{sec_analysis_bins}.
    \label{fig_bin}}
\end{figure}

In Figure\,\ref{fig_bin} we illustrate and compare our method of determining
$\Hlim$ with the previously established method of using the
exponential-fitting method. In the example, we perform both $\Hlim$ estimation
methods at a single location (radial bin) containing 2041 observed objects with
$a \in (2.693, 2.695]$~(au).  

For this example, we find the peak of the distribution (the most populated $H$
bin) to be at $\Hlim = 17.1\pm 0.29$, where the uncertainty is the 1-$\sigma$
confidence interval. Regarding our bin widths in $H$, we note that some of the
reported magnitudes in the MPC data have precision of 0.5 magnitude, the
majority have precision of 0.1 magnitude, and some objects are reported to
better precision. We used an $H$ bin width of 0.25 magnitude, but found
consistent results when using bin widths of 0.5 to 0.05.  Users of our approach
are encouraged to consider bin widths appropriate for their data set.

For the exponential-fitting method, one must determine what range of magnitude
values are to be included in the fitting.  In this example we choose the range:
$H\in[12.5,16.75]$. The adoption of a somewhat different range of values could
be easily justified.  We have chosen a range that, while not being an
unreasonable range for the asteroid belt, 
shows the greatest disagreement between the two methods; this illustrates 
that the two
methods result in similar $\Hlim$ for this sample.  The data points used in
fitting are shown in black.  We perform Bayesian linear regression
\citep{Kelly2007} on the data and the resulting best fit is shown as a blue
line, with the corresponding confidence interval shown in lighter blue.

It is evident that one might reasonably choose ``by eye" the location marked
with a red `x' (at magnitude 16.75) as the completion limit because it is where
the data deviate visibly from the black line (the linear best-fit). As can be
seen in the figure, this point lies within the 3-$\sigma$ confidence limit of
the linear fit.  In this example, the exponential-fitting method and our method both
agree on $\Hlim\sim17$, but the exponential-fitting method requires subjective choices.

We expect our method to result in values of $\Hlim$ that are
systematically higher than those found by the exponential-fitting method.
In order to get an idea
of how much higher, we take 50 random semimajor axis bins (of width 0.002 au)
between 2.6 and 2.8\,au and measure the distance between the exponential fit and
our estimated $\Hlim$.  We find that the median distance is +0.24 magnitude
with a standard deviation of 0.09 magnitude.  This should be taken as a rather
rough estimate because it is not clear that the exponential-fitting method represents the
true completion limit value as the size distributions of minor planets are not
consistent with single power-laws \citep{Cellino1991, Jedicke1998, Yoshida2005}
and its results are impacted by choices of which data is to be included and
which excluded in the exponential fitting.

Our method assumes that the peak in the observed 
$H$ distribution represents $\Hlim$.  This assumption may not be valid if the
distribution is multimodal.  A local maximum in the intrinsic 
distribution 
could be confused with $\Hlim$ in the apparent distribution if it were significant enough to
exceed the population of objects observed at the true $\Hlim$.  While our method does not
assume that minor planet size-frequency distributions follow single power-laws,
and our method can be compatible with distributions that contain features or
are not monotonically increasing, it is important to note that our method may
return the wrong $\Hlim$ in the above case.  This may be important to consider
for some populations, 
e.g., the Kuiper Belt \citep{2013ApJ...764L...2S, 2019Sci...363..955S}.

Finally, it should be noted that our $\Hlim$ represents the magnitude
at which observational completeness begins falling faster than the population of
asteroids is increasing with decreasing size.  In observational surveys for small bodies in the outer solar system, \cite{Bannister2016} show that detection
efficiency can fall by 20\% over 2 magnitudes before falling off much more
quickly.

\subsection{Model fitting} \label{sec_model_fitting}

Our initial tests in fitting the measured $\Hlim$ as a function of $a$ found that a linear modeling
of $\Hlim$ could fit fairly well much of the data, but could not
simultaneously fit the inner and outer edges of the main belt.  This slight
curve seen in Figure\,\ref{fig_heatmap} (monotonically decreasing, concave
up) motivated us to find a non-linear function for our fitting, which we describe next.

We start with the premise that asteroids are observed at or near opposition with Earth-based telescopes detecting reflected sunlight. 
The flux of reflected sunlight from an asteroid of diameter $D$ and geometric albedo $p$ located at a heliocentric distance $a$ is 

\begin{equation}
 F = {pD^2L_\odot\over 64\pi a^2(a-1\,\au)^2}, 
\end{equation}
where $L_\odot$ is solar luminosity.
If $F_{\rm lim}$ is the smallest detectable flux, then in flux-limited observations, the diameter of the smallest asteroid that is detectable at heliocentric distance $a$ is
\begin{equation}\label{eq:Dlim}
D_{\rm lim} = \Big({64\pi F_{\rm lim}\over pL_\odot}\Big)^{1\over2} a(a-1\,\au).
\label{e:Dlim}\end{equation}

To convert $D_{lim}$ to $H_{lim}$, we use the standard definition of absolute magnitude
for asteroids, $H$, as the apparent visual magnitude\footnote{The astronomical
magnitude system measures brightness on a logarithmic scale, such that apparent
brightness magnitude is defined as $m = m_{0} -2.5 \log_{10} (F/F_{0})$, where
$m_{0}$ is a standard reference brightness and $F_{0}$ is a standard
normalization flux; see \citet{Shu1982} and \url{https://cneos.jpl.nasa.gov/glossary/h.html}} at distance 1 au from the Sun and 1 au from the observer (at zero phase angle), i.e., 
\begin{equation}
H = H_{0} - 2.5 \log_{10} {F_1\over F_{0}} ,
\label{e:H}\end{equation}
where $H_{0}$ is a standard reference magnitude and $F_{0}$ is a standard normalization flux; here $F_1$ is the flux of reflected sunlight from an asteroid imagined to be at heliocentric distance 1 au and geocentric distance 1 au (at zero phase angle from an Earth-based observer), i.e., 
\begin{equation}
F_1 = {p D^2 L_\odot \over 64 \pi (1~{\rm au})^4 }.
\label{e:F1}\end{equation}
Setting $D=D_{lim}$ in Eq.~\ref{e:F1} and using Eq.~\ref{e:H}, it follows that the absolute magnitude $H_{lim}$ of the faintest detectable asteroid is

\begin{equation}
H_{lim} 
  = -5 \log_{10} (a(a-1~au)) + C ,
\label{e:Hlima}\end{equation}
where the constant $C$ depends upon the unknown effective limiting flux of asteroid surveys to date, 

\begin{equation}
C = H_{0} - 2.5 \log_{10} {F_{lim}\over F_{0}} .
\label{e:C}\end{equation}

In deriving the model equations~(\ref{e:Hlima}--\ref{e:C}), we have neglected phase angle effects; this can introduce some imperfection in our model for objects observed far from opposition, such as those  at higher inclinations and at closer distances. Our model is necessarily approximate as it is meant to account for broad trends with heliocentric distance, from the Asteroid Belt to Jupiter Trojans; we do not consider near-Earth asteroids in our modeling or analysis.

We use Equation~\ref{e:Hlima} for model-fitting, with $C$ as our free
parameter.  In our model-fitting, we use semi-major axis as a proxy for the
effective heliocentric distance of an asteroid; this is a reasonable
approximation for our purpose because most asteroids have been observed at many
points in their orbits so their semi-major axis is representative of their
average heliocentric distance.  We note that this is a strong assumption which appears to hold well for the main belt; we discuss it in more detail in Section~\ref{sec_discussion}.  

Taking the values of $\Hlim$ (measured at each semi-major axis bin) as our
data, we fit to it Equation~\ref{e:Hlima} using a Markov chain
Monte Carlo (MCMC) algorithm.  We use the MCMC ensemble sampler implemented in
the open-source code \texttt{emcee}
\citep{Goodman2010,Foreman-Mackey2013}.  The posterior distribution of
our free parameter ($C$, see eq.~\ref{e:C}) is sampled using
50 MCMC walkers.  These walkers are initialized as a random uniform
distribution $\sim$U(-50,50).  We find autocorrelation lengths (a measure of
the number of steps needed for modeling convergence) for our parameter from 8
to 17 steps (depending on region, see table~\ref{tab_fitting}).  From this, and in order to be conservative, we throw away the
first 500 steps (i.e.\ our ``burn in'' period) of each walker in order to end up with
$\sim$10,000 samples of our posterior distribution.
We use the posterior distribution generated by our MCMC fitting to find the
68\% confidence interval around $C$.

To estimate the uncertainty of the computed $\Hlim$, consider a
hypothetical bin with two objects; for such a bin, either $H_{min}$ or
$H_{max}$ could be taken to be $\Hlim$.  Whichever one is chosen, the
uncertainty should take account of the other value because the probability is
the same for each. We then adopt for the uncertainty the distance between the
two magnitudes. This value respects and includes both $H_{min}$ and $H_{max}$
within its bounds, and is connected to the bin population's distribution. Going
beyond a sample of two objects, because we expect that the uncertainty would
scale with sample size as in Poisson statistics, we divide by the square root
of the number of asteroids in the bin.  This approach estimates the measurement
error as $(H_{max}-H_{min})/\sqrt{n}$, where $H_{max}$ and $H_{min}$ are the
largest and smallest observed $H$ values within the considered region being
modeled and $n$ is the sample size in that region.

We have checked the effect of semi-major axis bin size on the measured value of $\Hlim$ and on the
model fitting results.  Semi-major axis bin widths of 0.01, 0.05 and 0.002 au
were tested, and we find that the resulting $\Hlim$ remained unchanged within
the uncertainties.  We adopted the bin width of 0.002 au for the model fitting,
but use the larger bin width of 0.01 au in Figure ~\ref{fig_fitting} and
Figure~\ref{fig:multi_fitting} for legibility reasons.

\subsection{Results}

\begin{figure}
    \plotone{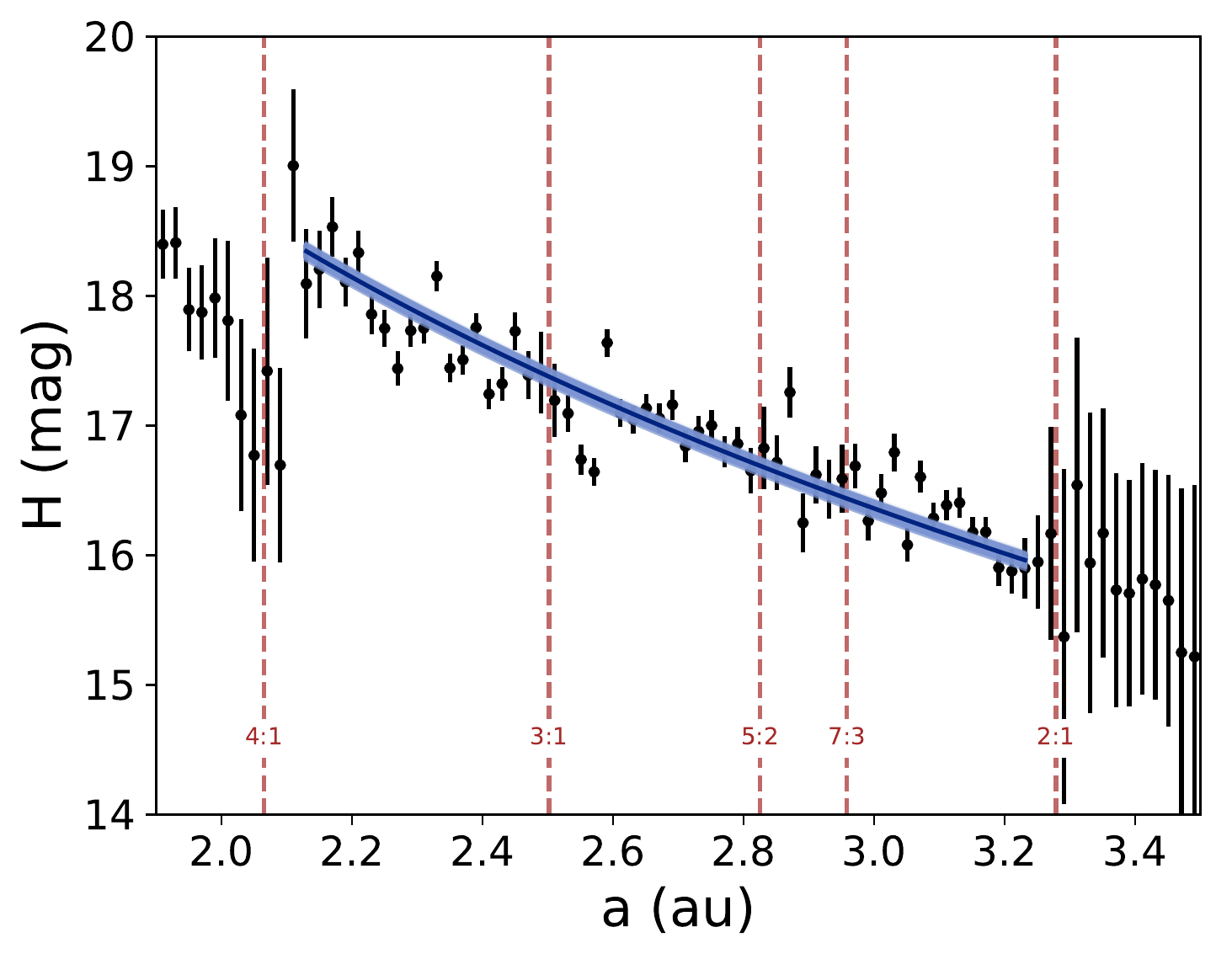}
    \caption{
    The completion limit, $H_{lim}$, at each semi-major axis bin is
    shown {as the black points; to make these legible, we used a larger semi-major axis bin size 0.01 au}.   
    {The analytical model (Equation~\ref{e:Hlima}) fit to the
    measured values of $\Hlim$ (at high resolution in $a$, with bin size 0.002 au) is shown in blue, and its 68\% credible interval
    is shown in light-blue.} 
     Dashed vertical lines (in red) denote a selection of mean motion
resonances with Jupiter.  \label{fig_fitting} }
\end{figure}

\begin{figure*}
\plotone{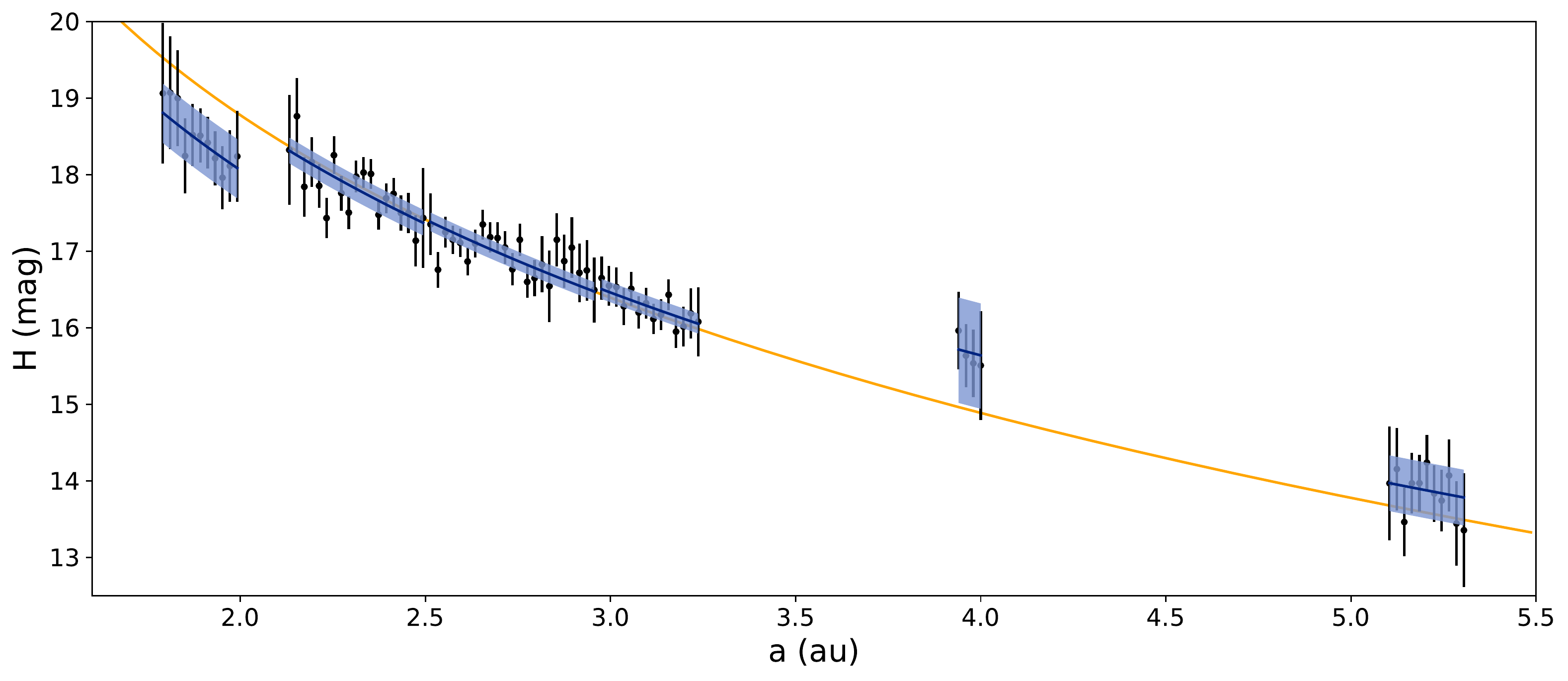}
\caption{
    The results of fitting our model to individual regions of asteroids in the semi-major axis range 1.78--5.30 au.
    Regions modelled are listed in Table \ref{tab_fitting}.
    Black dots indicate values of $\Hlim$ computed in semi-major axis bins of width $0.002$~au.  In this figure, we plot
    larger semi-major axis bin widths to make the figure more legible, see the end of 
    Section~\ref{sec_model_fitting} for the bin widths used in the actual model fitting.
    Black vertical lines show the uncertainty in $\Hlim$.  The best fit
    for each region is shown in blue, with
    its associated 68\% credible region in light-blue shading.  The main belt fit (see
    figure~\ref{fig_fitting}) is extended through the entire plot (orange line)
    to highlight deviations of other regions relative to the main belt model.
    }
    \label{fig:multi_fitting}
\end{figure*}

Figure~\ref{fig_fitting} shows the results of fitting our
model to the entire main asteroid belt (2.12 -- 3.25 au). 
{We also fit several subregions} independently, and we show the individual completion limit {model} fits for these regions in
Figure~\ref{fig:multi_fitting}.  
Table~\ref{tab_fitting} lists these
regions, their inner and outer edges as we have defined them, and the most likely
value for the parameter $C$ found from our fitting with the associated 68\%
credible region.

\begin{table}
\renewcommand{\thetable}{\arabic{table}}
\caption{Model fitting results} \label{tab_fitting}
\begin{tabular}{lrrrll}
\tablewidth{0pt}
\hline
\hline
    Region   & inner & outer & $n$    & $C$   & $\Hlim$ \tablenotemark{a}\\
             & (au)  & (au)  &        & (mag) & (mag)   \\
\hline
   Hungarias & 1.780 & 2.000 & 23527  & $19.57 \pm 0.14$  & 18.38   \\
   main belt & 2.120 & 3.250 & 783022 & $20.28 \pm 0.032$ & 17.01   \\
  inner belt & 2.120 & 2.500 & 246192 & $20.23 \pm 0.058$ & 17.76   \\
 middle belt & 2.500 & 2.960 & 328598 & $20.28 \pm 0.044$ & 17.01   \\
  outer belt & 2.960 & 3.250 & 208232 & $20.35 \pm 0.047$ & 16.28   \\
      Hildas & 3.920 & 4.004 & 3656   & $21.04 \pm 0.24$  & 15.69   \\
     Trojans & 5.095 & 5.319 & 6784   & $20.58 \pm 0.13$  & 13.88   \\
\hline
\end{tabular}
    \tablenotetext{a}{Median $\Hlim$ for the given region.}
\tablecomments{Uncertainties reported for $C$ are 68\% confidence intervals.}
\end{table}

{As can be seen in Figure~\ref{fig_fitting}, the model is a good fit in the range of the main asteroid belt where the sample is most populous, $2.12\,\au <a<3.25$~au, although we do observe statistically significant deviations of a few individual semi-major axis bins.}
Across the main belt, our model finds a completion limit of
$\Hlim=18.4$ at the innermost edge at $a=2.12$~au, and $\Hlim=15.9$ at the outermost edge at $a=3.25$~au. 
Regions within the main belt (inner, middle, outer), when fit individually,
result in model parameters that agree closely (within 3-sigma) with the entire
main belt fitting.

However, when we compare the regions beyond the main belt, we find $C$ to vary
by 1.5 magnitudes.  This is evident also when, as a fiducial reference, we
extend the completion limit model for the main belt to other regions, we find
that $\Hlim$ is over-predicted for the Hungaria objects, and under-predicted
for the Hilda and Trojan objects.  These differences likely arise from
deviations from the main belt in the size, and
eccentricity/inclination distributions of these populations (see Section \ref{sec_discussion}).
Differences in survey coverage of these three population groups likely also
contribute to these deviations.

\section{Discussion \& Summary} \label{sec_discussion}

Knowing the completion limit of the minor planets' observational data is
necessary in order to define an observationally complete sample before
performing statistical analyses or testing theoretical models of the dynamical
and physical evolution of asteroids.  For example, before one can search for
features in their size frequency distribution, the trend imparted on the data
from this observational bias must be removed.

The use of a single $\Hlim$ for the entire asteroid belt (or for sub-regions)
may be appropriate for determining an observationally complete sample if a
sufficiently conservative choice is made.  However, for some applications, this may lead to a smaller
sample size even when a significantly larger one is available, thereby
``leaving data on the table" and limiting the statistical confidence of a
study. Moreover, in a situation where one hopes to characterize a subtle trend
throughout the asteroid belt (or a sub-region such as the inner asteroid belt),
$\Hlim$ would need to be known not just as one value for
the region, but throughout the region as a function of $a$.  This requires
estimating $\Hlim$ many times at high resolution in $a$.

Previous methods of determining $\Hlim$ (i.e., by eye, or by the
exponential-fitting method) present several drawbacks.  Determining $\Hlim$ by eye has an obvious
problem with reproducibility; it is also impractical to do more than a few 10s
of times for any extended portion of the asteroid belt.  The exponential-fitting method
also presents several issues to consider.  To determine where the
size-distribution of a sample of objects deviates from a power-law function
requires assumptions that make reproducibility difficult: first, a range of
H-values needs to be identified for inclusion in the exponential fitting, and
second, one must choose the amount of deviation from an exponential which is large
enough to identify the limiting magnitude.  Additionally, the data samples
being fit may not be well represented by single power-law size distributions
\citep{Cellino1991, Jedicke1998, Yoshida2005}.

We presented here our method of modeling $\Hlim$ which solves several of these
problems.

First, by identifying the most populated $H$ values throughout a
range of $a$, our approach is agnostic about the underlying size distribution.
Second, there are fewer decisions that need to be made when compared to the
exponential-fitting method.  This gives our approach the advantage that it
is more easily reproducible (although the exponential-fitting method
is reproducible in principle, the specific decisions in implementing it are
usually not specified in the published literature).

Third, it is easily
implemented in flexible ways: the $\Hlim$ model can be computed as a single
value for an extended region, or one can compute $\Hlim(a)$ as a function of
semi-major axis with (user-determined) high resolution in $a$.
In addition,
because our approach is relatively hands-off, it scales to large data sets.
This would be especially important in the coming decade as the the Vera C.~Rubin Observatory's Legacy Survey of Space and Time (LSST) 
increases the observational sample of asteroids by an order of
magnitude \citep{Ivezic2019}. 

As described in Section~\ref{sec_analysis}, we have measured the empirical
completion limit of minor planets (as a function of $a$) within well populated
regions between 1.78~au and 5.30~au.  We then use the results to fit to our
physically motivated completion limit model
(Equation~\ref{e:Hlima}).  Table~\ref{tab_fitting} reports the
regions, the number of objects in each region ($n$), the most likely value of
our fit parameter $C$, and median $\Hlim$ for the given region, and Figure~\ref{fig:multi_fitting} plots the best-fit models.

As noted above,
our model finds a completion limit of
$\Hlim=18.4$ at the innermost edge of the main belt, at $a=2.12$~au, and $\Hlim=15.9$ at the outermost edge at $a=3.25$~au.
This lowest value found is larger than many of the completion limit values
adopted in recent studies, suggesting that those studies adopted somewhat
conservative completion limits.  
(Note, however, that $\Hlim$ is steadily increasing over time as surveys become wider and deeper, consequently it is not straightforward to directly compare our $\Hlim$ with its value in previous studies.)
If we were to make the conservative choice of adopting a single value, $\Hlim=15.9$
for the entire main belt, which is the approach taken in many previous studies, then our
sample size would be 201,701. 
However, if we instead define our sample as $H<\Hlim(a)$,   
the sample size is 465,837 objects, a factor of 2.31 larger. A larger sample size enables correspondingly greater statistical confidence.

The model fit, $\Hlim(a)$  (Eq.~\ref{e:Hlima}), for the main belt accounts quite well for the broad trends with heliocentric distance in the main belt (see Fig.~\ref{fig_fitting}). When we extend this model to regions outside of the main belt we find that the measured $\Hlim(a)$ deviates significantly from the main belt model.  In Figure~\ref{fig:multi_fitting} one can see that our main belt model over-predicts $\Hlim(a)$ for Hungaria group and under-predicts it for the Hilda and Trojan groups.  These discrepancies may result from differences in the distribution of sizes, eccentricities and/or inclinations amongst these regions; the causes of the discrepancies may be different for different regions. For the Hungarias, their smaller heliocentric distances and higher inclinations (16$^\circ$--34$^\circ$) are factors that reduce their detectability at low phase angles, qualitatively consistent with their observed discrepancy relative to the main belt model prediction in Figure~\ref{fig:multi_fitting}.  For the Hildas and Trojans, the explanation may lie in the significant dispersion of their eccentricities. It is conceivable that  observational data of these objects is obtained more often when they are near their perihelion distance, so that our estimate of $\Hlim$ for these groups may be more representative of their perihelion distance rather than their semi-major axis. This explanation is qualitatively consistent with their observed discrepancy relative to the main belt model prediction in Figure~\ref{fig:multi_fitting}: the $\Hlim$ values for these groups should be associated with systematically smaller heliocentric distance than their values of semi-major axis, $a$.  The cause of these discrepancies is worth exploring quantitatively in the future.

It is our recommendation that future works consider the completion limit a
function of semi-major axis, and we offer our parameterized model (and single
values, for individual regions) for that purpose. $H_{lim}$
-- which will increase over time as surveys get deeper and wider --
can be recalculated for any set of objects using the
MPCORB database and our code available at
\url{http://github.com/equant/Asteroids/}.

\acknowledgments

We thank two reviewers (Jean-Marc Petit and anonymous) for comments that helped to improve this paper. We acknowledge funding from NASA Nexus for Exoplanet System Science (NExSS; grant NNX15AD94G) and the Marshall Foundation of Tucson, AZ.

This research has made use of data and/or services provided by the International Astronomical Union's Minor Planet Center.

%



\software{astropy,  
            emcee \citep{Foreman-Mackey2013},
            linmix
          }




\bibliography{references-autogenerated-ads,references-custom}

\end{document}